\documentclass{PoS}

\usepackage{amsmath,subfigure,latexsym,amssymb}

\newcommand{\be}{\begin{equation}}
\newcommand{\ee}{\end{equation}}

\newcommand{\bea}{\begin{eqnarray}}
\newcommand{\eea}{\end{eqnarray}}

\newcommand{\R}{{\kern+.25em\sf{R}\kern-.78em\sf{I} \kern+.78em\kern-.25em}}
\newcommand{\RR}{{\kern+.25em\sf{R}\kern-.6em\sf{I} \kern+.6em\kern-.25em}}
\newcommand{\N}{{\kern+.25em\sf{N}\kern-.78em\sf{I} \kern+.78em\kern-.25em}}
\newcommand{\C}{{\kern+.25em\sf{C}\kern-.50em\sf{I} \kern+.50em\kern-.25em}}

\makeatletter
\@addtoreset{equation}{section}
\makeatother

\title{O(N) Models with Topological Lattice Actions
\thanks{We thank Janos Balog, Peter Weisz and Ulli Wolff for helpful 
communications. \newline
This work was supported in parts by the {\it Schweizerischer 
Nationalfonds} (SNF), and by the Mexican {\it Consejo Nacional de 
Ciencia y Tecnolog\'{\i}a} (CONACyT), project 155905/10 {\it F\'{\i}sica 
de Part\'{\i}culas por medio de Simulaciones Num\'{e}ricas} and 
through the scholarship 312631 for graduate studies, as well as 
DGAPA-UNAM.}}

\ShortTitle{Topological Lattice Actions}

\author{\speaker{Wolfgang Bietenholz}$^{\rm \ a}$, 
Michael B\"{o}gli$^{\rm \ b}$, Urs Gerber$^{\rm \ a}$,
Ferenc Niedermayer$^{\rm \ b,c}$, Michele Pepe$^{\rm \ d}$,
Fernando G.\ Rej\'{o}n-Barrera$^{\rm \ e}$ and
Uwe-Jens Wiese$^{\rm \ b}$ \\
\ \\
\ \\
$^{\rm \ a}$ Instituto de Ciencias Nucleares, Universidad Nacional 
Aut\'{o}noma de M\'{e}xico \\
~~~~A.P.\ 70-543, C.P.\ 04510 Distrito Federal, Mexico \vspace*{1mm} \\
\ $^{\rm b}$ Albert Einstein Center for Fundamental Physics \\
~~~~Institute for Theoretical Physics, Bern University \\
~~~~Sidlerstrasse 5, CH-3012 Bern, Switzerland  \vspace*{1mm} \\
\ $^{\rm c}$ Institute for Theoretical Physics -- HAS, 
E\"{o}tv\"{o}s University \\
~~~~P\'{a}zm\'{a}ny s\'{e}t\'{a}ny 1/a, 1117 Budapest, Hungary 
\vspace*{1mm} \\
\ $^{\rm d}$ INFN, Sezione di Milano-Bicocca, Edificio U2 \\
~~~~Piazza della Scienza 3, 20126 Milano, Italy  \vspace*{1mm} \\
\ $^{\rm e}$ Institute for Theoretical Physics, University of Amsterdam \\
~~~~Science Park 904, Postbus 94485, 1090 GL Amsterdam, The Netherlands \\
\ \\
E-mail: \email{wolbi@nucleares.unam.mx} \\ }

\abstract{A variety of lattice discretisations of continuum actions 
has been considered, usually requiring the correct classical 
continuum limit. Here we discuss ``weird'' lattice formulations
without that property, namely lattice actions that are invariant under 
most continuous deformations of the field configuration, 
in one version even without any coupling constants. 
It turns out that universality is powerful enough to 
still provide the correct quantum continuum limit, despite the absence 
of a classical limit, or a perturbative expansion. We demonstrate this 
for a set of O(N) models (or non-linear $\sigma$-models). 
Amazingly, such ``weird'' lattice actions are not only in the right
universality class, but some of them even have practical benefits, 
in particular an excellent scaling behaviour. \\ \ \\ }

\FullConference{31st International Symposium on Lattice Field Theory 
- LATTICE 2013\\
		July 29 - August 3, 2013\\
		Mainz, Germany}

\begin{document}

\section{Topological Lattice Actions}

Lattice field theory usually starts by discretising some continuum
Lagrangian, such as\footnote{In this article, we do not consider 
gauge fields.}
\be
{\cal L}(\Phi (x), \partial_{\mu} \Phi (x)) \ \longrightarrow \
{\cal L}_{\rm lat}(\Phi_{x}, %\frac{1}{a} 
[\Phi_{x + a \hat \mu} - \Phi_{x}] /a ) 
\ .
\ee
The couplings of a discrete derivative may also be spread somewhat 
beyond nearest neighbour sites but the continuum extrapolation of
physical quantities coincides.
% for a variety of lattice formulations of this kind. 
This is due to %the powerful property of 
{\em universality:} the universality class is determined by
the dimension of space-time and by the symmetries of the order
parameter fields. A condition is locality, {\it i.e.}\ the couplings
should decay at least exponentially with the distance, and it is
popular to tacitly assume that also the classical continuum limit
should reproduce the continuum Lagrangian, {\it e.g.}\
$ ^{\lim}_{a \to 0} [\Phi_{x + a \hat \mu} - \Phi_{x}] / a 
= \partial_{\mu} \Phi (x)$. 

Here we investigate counter-examples to the last assumption, namely 
lattice actions which do not have any classical limit. Thus we
are probing how far universality really reaches. It turns out
that the quantum continuum limit may still be correct,
and --- surprisingly --- such highly unconventional lattice
formulations even have practical virtues.

We study O(N) models, with classical spins of unit length attached to %on
each lattice site,
\be
\vec e_{x} = (e_{x}^{1}, \dots ,e_{x}^{N}) \ , \quad
| \vec e_{x} | = 1 \quad \forall \ x = n a \ , \quad n \in %\Z^{d} \ .
\mathbb{Z}^{d} \ .
\ee
We consider the dimensions $d =1$ and $2$, and $N = 2$
(XY model, relevant {\it e.g.}\ for superfluid $^{4}$He films)
and $N=3$ (classical Heisenberg model, asymptotically free,
describes ferromagnets). For $N = d+1$, periodic boundary 
conditions imply that the configurations occur in topological 
sectors (we employ the geometric definition of the topological 
charge of lattice configurations). % \\

The simplest and most radical topological action is the
{\em constraint action,} which just restricts the angles
between all pairs of nearest neighbour spins by an upper
bound $\delta$,\footnote{O(N) model simulations with
such a constraint have a pre-history, which includes 
Refs.\ \cite{prehist}.}
\be  \label{conact}
S [ \vec e \, ] = \sum_{\langle x,y \rangle} s (\vec e_{x}, \vec e_{y})
\ , \qquad s (\vec e_{x}, \vec e_{y}) = \left\{ \begin{array}{ccccc}
0 &&&& \vec e_{x} \cdot \vec e_{y} > \cos \delta \\
+ \infty &&&& {\rm otherwise} \end{array} \right. \quad .
\ee
Most small deformations of a configuration (those within the 
allowed set) do not cost any action; this characterises
{\em topological lattice actions.} All allowed configurations
have the same action $S = 0$; due to this enormous degeneracy,
there is no classical limit, nor a perturbative expansion.

For models with topological charges 
$Q = \sum_{\langle x,y,\dots \rangle} q_{\langle x,y,\dots \rangle}$
(where $q$ is the topological charge density), we
also consider the {\em $Q$ suppressing action}
\be
S [ \vec e \, ] = \lambda \sum_{\langle x,y,\dots \rangle} 
| q_{\langle x,y,\dots \rangle} | \ , \quad \lambda > 0 \ .
\ee
The 2d XY model does not have topological sectors, but each
plaquette carries a vortex number $v_{\Box} \in \{ 0, \pm 1 \}$,
which can be suppressed analogously,
$S [ \vec e \, ] = \lambda \sum_{\Box} | v_{\Box} |\,$.

We are going to consider constraint actions, $Q$ (or vortex)
suppressing actions, and combinations. All these are topological
lattice actions, since $S [ \vec e \, ]$ is invariant under
most small deformations of the configuration $[ \vec e \, ]$
(in contrast to lattice actions with discrete derivative terms).

\section{The Quantum Rotor}
\vspace{-1mm}

The 1d XY model describes a quantum mechanical particle moving freely 
on a circle, with the continuum action
$S [ \varphi ] = \frac{I}{2} \int_{0}^{\beta} dt \,
\dot \varphi (t)^{2}$ ($\varphi (t)$: angle, $I$ moment of inertia).
With periodicity, $\varphi (\beta ) = \varphi (0)~{\rm mod}~2 \pi$,
this is the simplest model with topological sectors.

For the constraint action (the $Q$ suppressing action)
the continuum limit is attained as $\delta \to 0$ 
($\lambda \to \infty$). The following table displays the
asymptotic behaviour for two scaling terms \cite{topact}.
\begin{table}[h!]
\renewcommand{\arraystretch}{1.4}
\centering
\begin{tabular}{|c||c|c|c|}
\hline
scaling term & continuum & constraint action & $Q$ suppressing action \\
\hline
\hline
$\frac{E_{2}-E_{0}}{E_{1}-E_{0}}$ & $4$ & 
$ 4 ( 1 + \frac{3}{5} \frac{a}{\xi} + \dots \ ) $ & 
$ 4 ( 1 - \frac{3}{2} \frac{a}{\xi} + \dots \ ) $ \\
\hline
$ \chi_{t} \, \xi = \frac{\langle Q^{2} \rangle }{\beta (E_{1}-E_{0})} $ 
& $\frac{1}{2 \pi^{2}}$ &  $\frac{1}{2 \pi^{2}} 
( 1 - \frac{1}{5} \frac{a}{\xi} + \dots \ ) $ 
& $\frac{1}{2 \pi^{2}}
( 1 + \frac{1}{2} \frac{a}{\xi} + \dots \ ) $ 
\\ \hline
\end{tabular}
\end{table}

The topological actions have {\em linear} scaling artifacts, 
which are unusual in scalar theories. However,
%more importantly, 
the continuum values are consistently
reproduced for $a/\xi \to 0$ ($\xi$: correlation length,
$\chi_{t}$: topological susceptibility).
This is non-trivial; we do observe a facet of universality even in 
quantum mechanics, although universality is assumed 
to hold 
only in field theory, $d \geq 2$.

\vspace*{-1mm}
\section{The 2d O(3) Model}

The 2d O(3) model with coupling $g$ (and periodic boundaries)
has the continuum functionals
\be
S [ \vec e \, ] = \frac{1}{2g^{2}} \int d^{2}x \ \partial_{\mu} 
\vec e \cdot \partial_{\mu} \vec e \ , \quad
Q [ \vec e \, ] = \frac{1}{8 \pi} \int d^{2}x \ \epsilon_{\mu \nu} \,
\vec e \cdot ( \partial_{\mu} \vec e \times \partial_{\nu} \vec e ) \in
\mathbb{Z} \ ,
\ee
which obey the Schwarz inequality 
$S [ \vec e \, ] \geq \frac{4 \pi}{g^{2}} \, | Q [ \vec e \, ] |$
for each configuration.

On the {\em lattice,} the geometric topological charge takes the form
$Q[ \vec e \, ] = \frac{1}{4 \pi} \sum_{\langle x,y,z \rangle} A_{x,y,z}$,
where $x,y,z$ are corners of a triangle (half a plaquette),
%\footnote{The optimal decomposition
%is shown graphically in Ref.\ \cite{topact}.}, 
and $A_{x,y,z}$ is the oriented area of the (minimal) spherical 
triangle spanned by $\vec e_{x}, \vec e_{y}, \, \vec e_{z} \, $.
%Regarding the lattice actions, 
We consider the standard lattice action,
$ S[ \vec e \, ] = - \frac{1}{g^{2}} \sum_{x,\mu}
\vec e_{x}\cdot \nolinebreak \vec e_{x + a \hat \mu}$, the constraint action
(\ref{conact}) and the $Q$ suppressing action
$ S[ \vec e \, ] = \lambda \sum_{\langle x,y,z \rangle} |A_{x,y,z}|$.

\begin{figure}
\vspace*{-8mm}
\begin{center}
\includegraphics[angle=270,width=.5\linewidth]{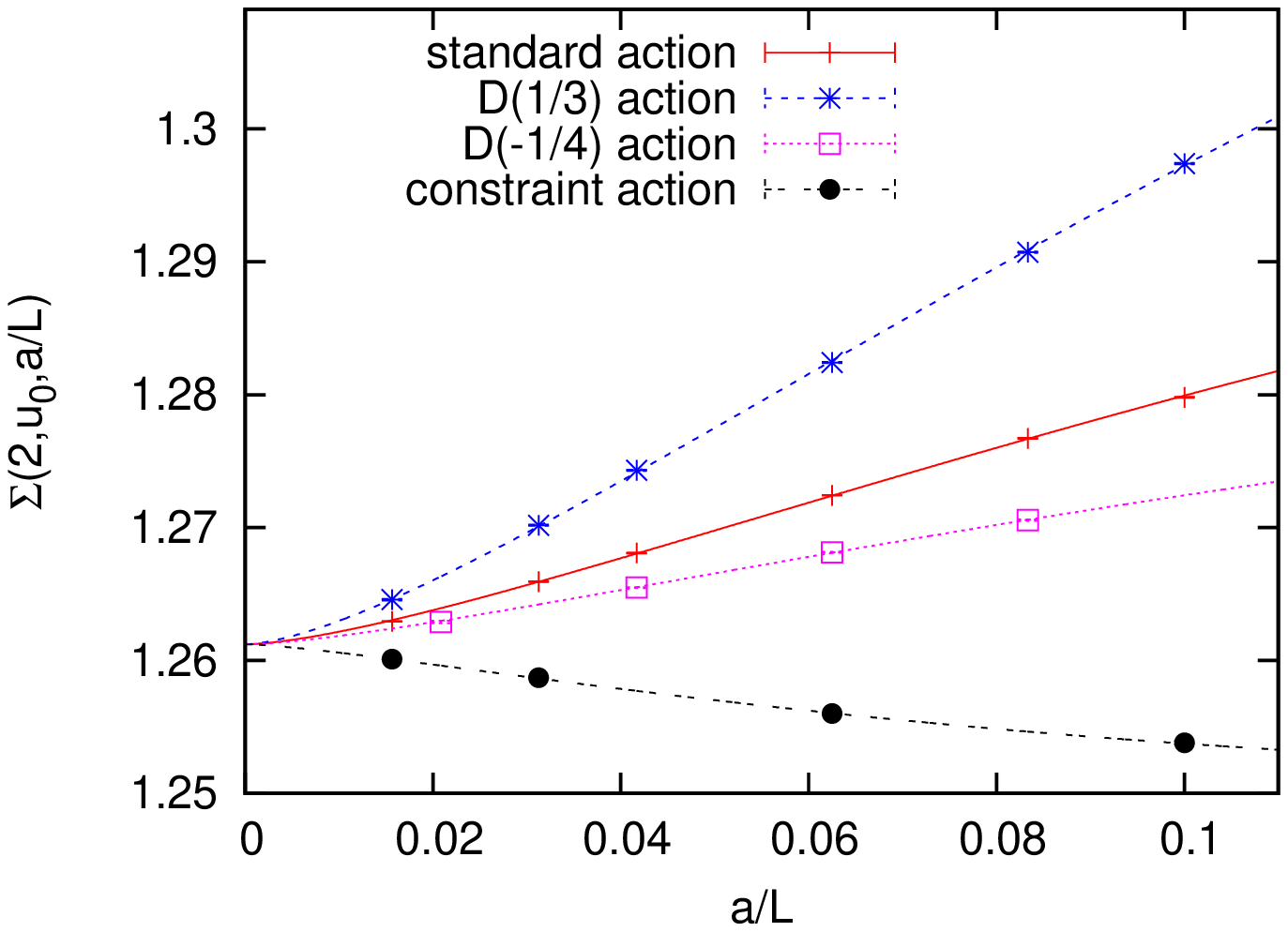}
\hspace*{-3mm}
\includegraphics[angle=270,width=.5\linewidth]{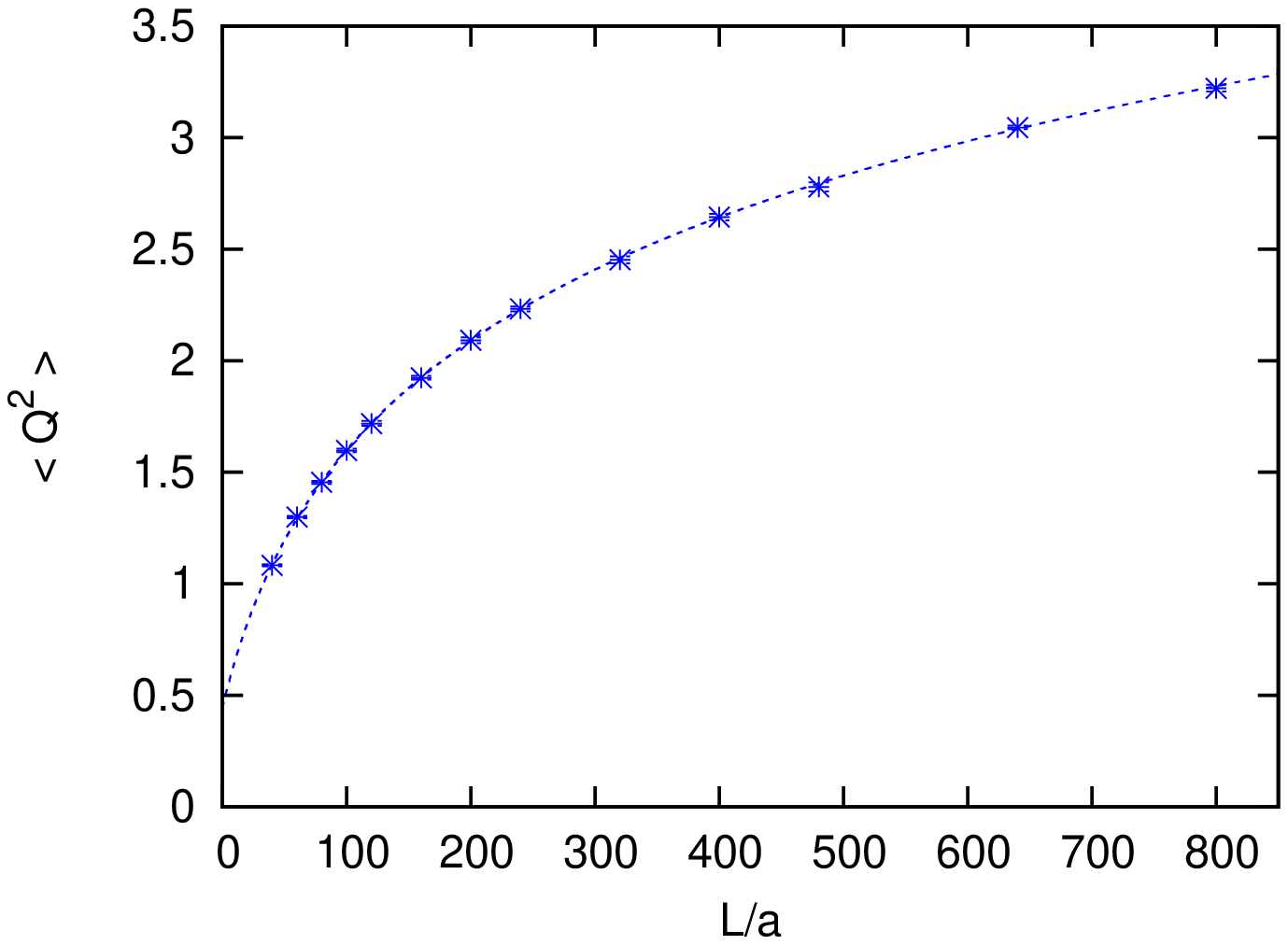}
\end{center}
\vspace*{-5mm}
\caption{On the left: The step-2 SSF for the 2d O(3) model with $u_{0} = 
1.0595$, for four lattice actions. The constraint action has the
correct continuum extrapolation, and the best scaling behaviour.
On the right: The ``scaling term'' $16 \chi_{t} \xi_{2}^{2} = 
\langle Q^{2} \rangle$ diverges only {\em logarithmically} 
for the constraint action, although dislocations
are not suppressed by any Boltzmann factor.} 
\vspace*{-3mm}
\label{SSF2dO3}
\end{figure}

As a scaling test we evaluated, on $L \times 10 \, L$ lattices,
the step-2 Step Scaling Function (SSF) \cite{LWW} $\sigma (2, u_{0}) = 
2 L/\xi (2L)$, with $u_{0} = L/\xi (L)$. The continuum value 
$\sigma (2, u_{0} = 1.0595) = 1.26121$ \cite{BNW} must
be reproduced in the continuum extrapolation of simulation
results with any lattice action in this universality class.
For the constraint action, we performed precise numerical 
measurements with the Wolff cluster algorithm. Figure \ref{SSF2dO3} 
(left) shows that the result is consistent with the continuum value.
The data can be fitted to the same ansatz as the standard and 
modified actions,\footnote{However, a recent discussion of O(N) 
models at large $N$ suggests that the power of the leading logarithmic 
term might differ for the constraint action \cite{drastic}.}
\be
\Sigma (2, u_{0} , a/L) = \sigma (2, u_{0})
+ \frac{a^{2}}{L^{2}} \Big( c_{1} \ln^{3}\frac{a}{L} +
c_{2} \ln^{2}\frac{a}{L} + \dots \Big) \ .
\ee
Amazingly, the constraint action even
%does not only have the correct continuum limit, but it 
scales better than all the conventional actions included in Figure 
\ref{SSF2dO3} (for them we took the data form Ref.\ \cite{BNW}).
\noindent

It has been predicted long ago that the ``scaling term''
$\chi_{t} \xi^{2}$ diverges in the continuum limit of this 
model ($\chi_{t} = \langle Q^{2} \rangle /V$). A semi-classical
argument refers to small topological dislocations, which are
insufficiently suppressed \cite{ML82}; that suggests a power divergence,
$\chi_{t} \xi^{2} \propto (\xi /a)^{p}$, $p \approx 0.9$. Simulations 
with a (truncated) classically perfect action --- which eliminates 
dislocations --- revealed a logarithmic divergence \cite{BBHN}. 

The opposite extreme is the constraint action, which (for the 
nearly critical $\delta$-angles of interest) 
allows for dislocations without any action cost. 
To investigate its behaviour, we fixed\footnote{$\xi_{2}$ 
is the second moment correlation length; it almost coincides 
with $\xi$, but is easier to measure. Its use was motivated
in particular by the $Q$ suppressing action, where we could
not apply the cluster algorithm.}
$L/\xi_{2}=4$ and measured $16 \chi_{t} \xi_{2}^{2} = \langle Q^{2} 
\rangle$ as a function of $L/a = 4 \xi_{2}/a$. The results for the 
constraint action (Figure \ref{SSF2dO3}, right) and for the $Q$ 
suppressing action still diverge only logarithmically \cite{topact}. 
%This power-free
%divergence is remarkable in particular for the constraint action.
%\begin{figure}
%\vspace*{-1cm}
%\begin{center}
%\hspace*{-9mm}
%\includegraphics[angle=270,width=.45\linewidth]{talk/figure9.ps}
%\hspace*{-12mm}
%\includegraphics[angle=270,width=.45\linewidth]{talk/figure10.ps}
%\end{center}
%\vspace*{-8mm}
%\caption{The ``scaling term'' $16 \chi_{t} \xi_{2}^{2}$ only diverges
%{\em logarithmic}, both for constraint action (left, dislocations 
%{\em not} suppressed)and for the $Q$ suppressing action (right).}
%\vspace*{-4mm}
%\label{susdivlog}
%\end{figure}

Due to this divergence, this model is sometimes considered ``sick'',
at least regarding topological properties. However, the correlation
of the topological density, $\langle q_{x} q_{y} \rangle$, does
have a regular continuum limit at non-zero separation \cite{CEPS}, 
$x \neq y$, as we confirmed for the topological actions \cite{topact}.

\vspace*{-2mm}
\section{The 2d XY Model}
\vspace*{-2mm}

Here we can express the spins as $\vec e_{x} = (\cos \varphi_{x}, 
\sin \varphi_{x} ) \in S^{1}$. For the relative angle between
nearest neighbour spins we define the mod operation such that
$
\Delta \varphi_{x,x + a \hat \mu} = ( \varphi_{x} - \varphi_{x + a \hat \mu})
\ {\rm mod} \ 2 \pi \in ( - \pi, \pi ] \ .
$
A plaquette $\Box$ with the corners $x_{1}, \, x_{2}, \, x_{3}, \, x_{4}$
has the {\em vortex number}
\be
v_{\Box} = \frac{1}{2 \pi} (\Delta \varphi_{x_{1},x_{2}} +
\Delta \varphi_{x_{2},x_{3}} + \Delta \varphi_{x_{3},x_{4}} +
\Delta \varphi_{x_{4},x_{1}} ) \in \{ 0, \pm 1 \} \ . 
\ee
Periodic boundary conditions imply $\sum_{\Box} v_{\Box} = 0$.

For the standard action $S [ \vec e \, ] = \beta \sum_{x,\mu}
(1 - \vec e_{x} \cdot \vec e_{x + a \hat \mu})$ there is a well-known
Berezinskii-Koster\-litz-Thouless (BKT) essential phase transition
(of infinite order) at $1/\beta_{\rm c} = 1.1199(1)$ \cite{MHas}, 
where the correlation length diverges exponentially,
\be
\xi ( \beta \lesssim \beta_{\rm c} ) \propto \exp ({\rm constant}/
\sqrt{\beta_{\rm c} - \beta} ) \ .
\ee

The established picture describes this transition by the vortex
dynamics: at $\beta > \beta_{\rm c}$ they occur (mostly)
in tightly bound vortex--anti-vortex pairs, which leads to a 
{\em massless phase.} As we decrease $\beta$ below $\beta_{\rm c}$ 
these pairs unbind and ``disorder'' the system, so we enter
the {\em massive phase.} The value of $\beta_{\rm c}$ has been
estimated based on the action cost for isolated vortices
(or anti-vortices) \cite{Kos74}. 

\begin{figure}
\vspace*{-7mm}
\begin{center}
%\hspace*{5cm} 
\includegraphics[angle=0,width=.49\linewidth]{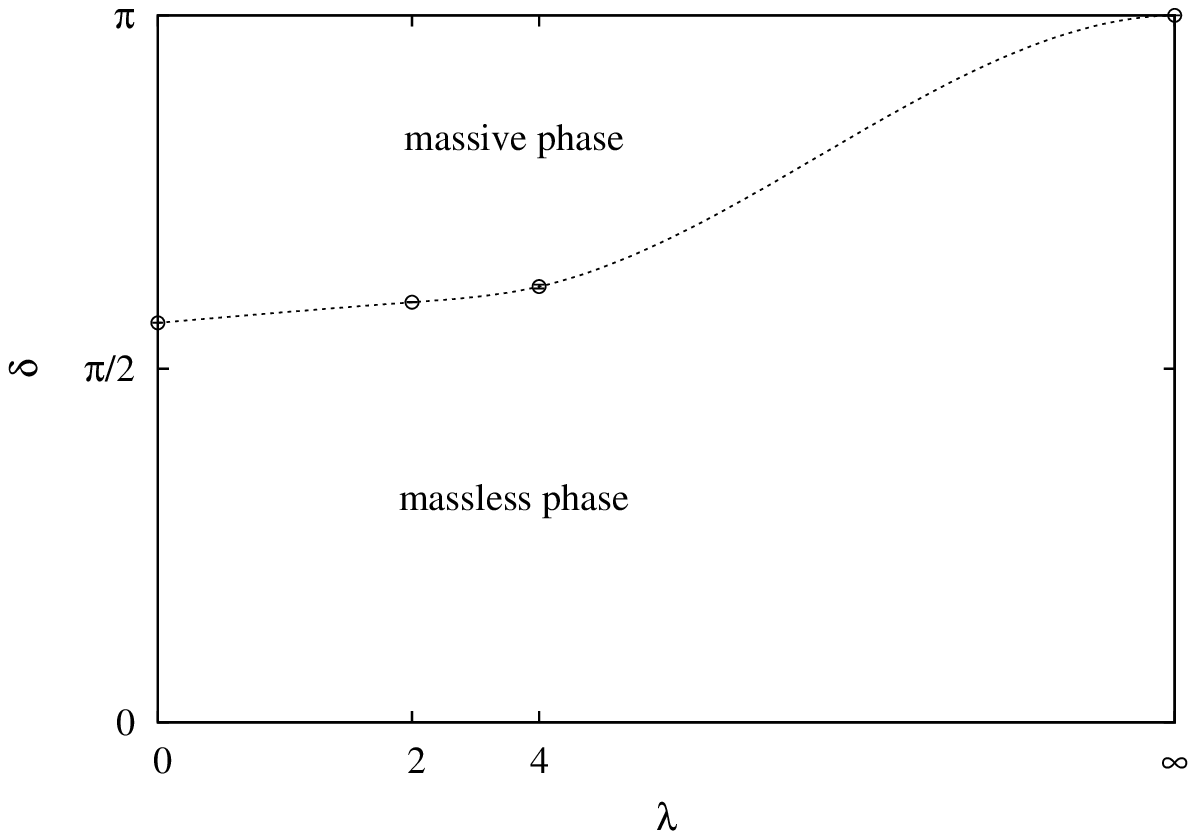}
\includegraphics[angle=0,width=.49\linewidth]{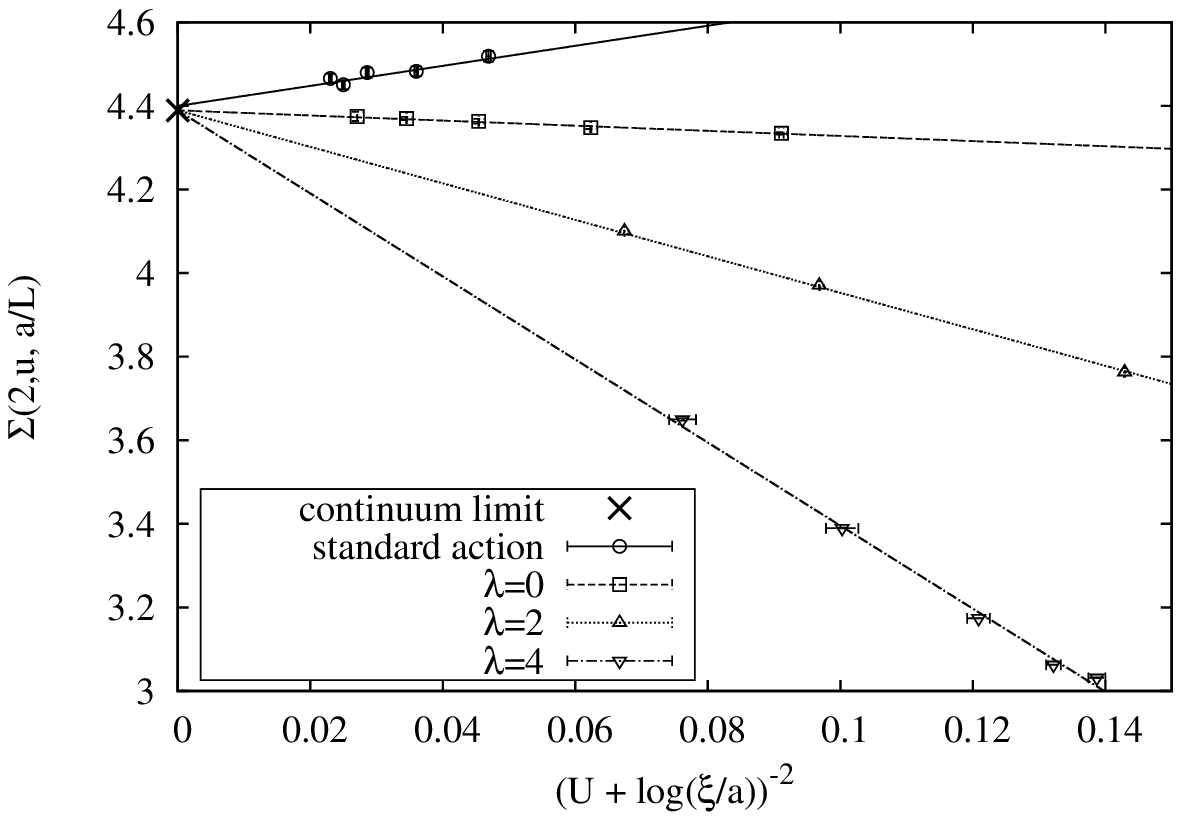}
\end{center}
\vspace*{-7mm}
\caption{On the left: Schematic phase diagram for the 2d XY model
with topological lattice actions. On the right: step-2 SSF for the
standard action and three topological lattice actions, which are
all compatible with the correct continuum extrapolation (for a suitable 
constant $U$; the continuum value $\sigma (2,3.0038)$ is included
in the fit). Also here the constraint action $(\lambda =0)$ has a 
formidable scaling quality.} 
\vspace*{-3mm}
\label{phasediaSSF}
\end{figure}

For the constraint action ---  with 
$| \Delta \varphi_{x, x + a \hat \mu} | < \delta$  for all $x,\mu$ 
--- vortices are either excluded ($\delta \leq \pi/2$), or 
allowed without any action cost. We simulated the constraint action,
the vortex suppressing action $S = \lambda \sum_{\Box} |v_{\Box}|$
and combinations (for $\lambda >0$ we elaborated a new variant of 
the cluster algorithm). At fixed $\lambda = 0,\, 2$ or $4$ we 
observed massive/massless phase transitions at \cite{XYtopact} 
\be
\delta_{\rm c} (\lambda = 0) = 1.7752(6)\ , \quad
\delta_{\rm c} (\lambda = 2) = 1.8665(8)\ , \quad
\delta_{\rm c} (\lambda = 4) = 1.936(8)\ ,
\ee
where the correlation length exhibits a BKT type divergence,
\be
\xi (\delta \gtrsim \delta_{\rm c}) \propto \exp ({\rm constant}
/ \sqrt{\delta - \delta_{\rm c}} \, ) \ . 
\ee
The phase diagram is sketched in Figure \ref{phasediaSSF} (left). 
In the vicinity of the transition, vortices are present 
(they are ruled out for $\delta < \pi /2$ and $\lambda = + \infty$).
To probe the BKT behaviour further, we measured again 
the step-2 SSF, now referring to the continuum value
$\sigma (2, u = 2L/ \xi (L) = 3.0038) = 4 L / \xi (2 L) = 4.3895$, 
which had been confirmed for the standard action \cite{BKKW}.
Figure \ref{phasediaSSF} (right) shows those data,
and our results for three topological actions, which
are all compatible with a fit to the right continuum value. 
Again the constraint
action (with $\lambda =0$) has an excellent scaling behaviour.

\begin{figure}[h!]
\vspace*{-9mm}
\includegraphics[angle=270,width=.49\linewidth]{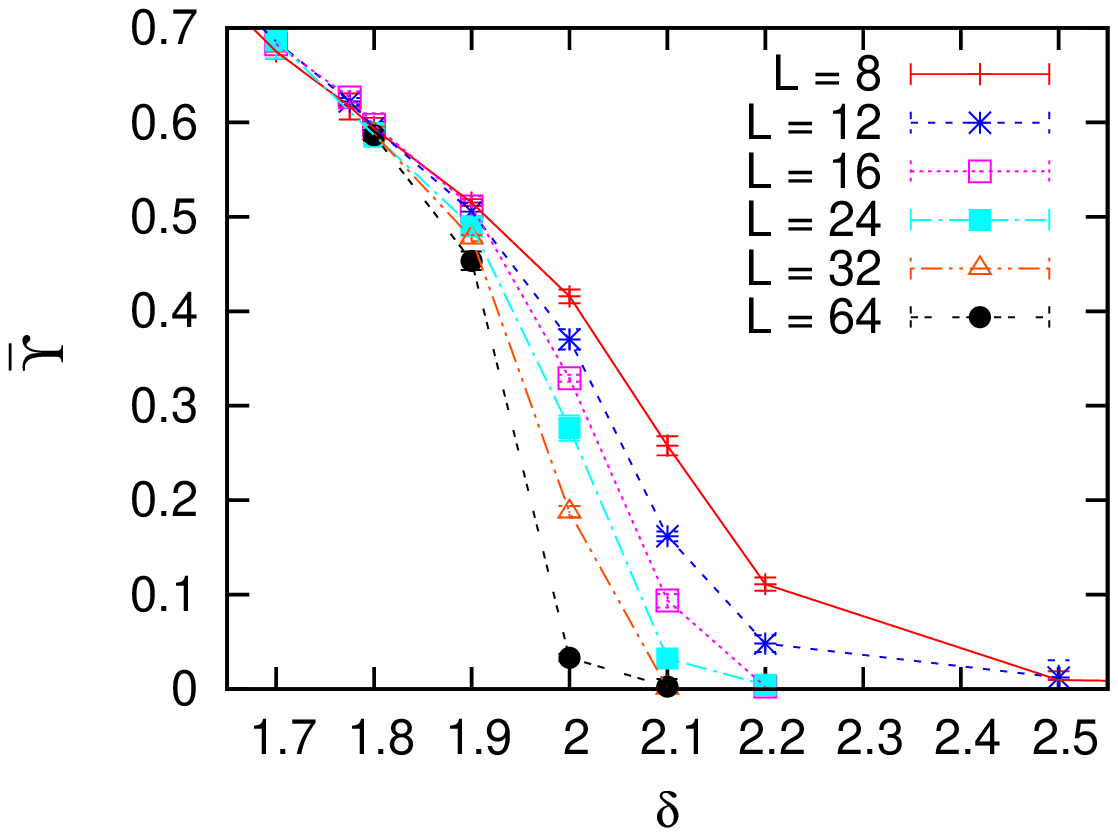}
\includegraphics[angle=270,width=.49\linewidth]{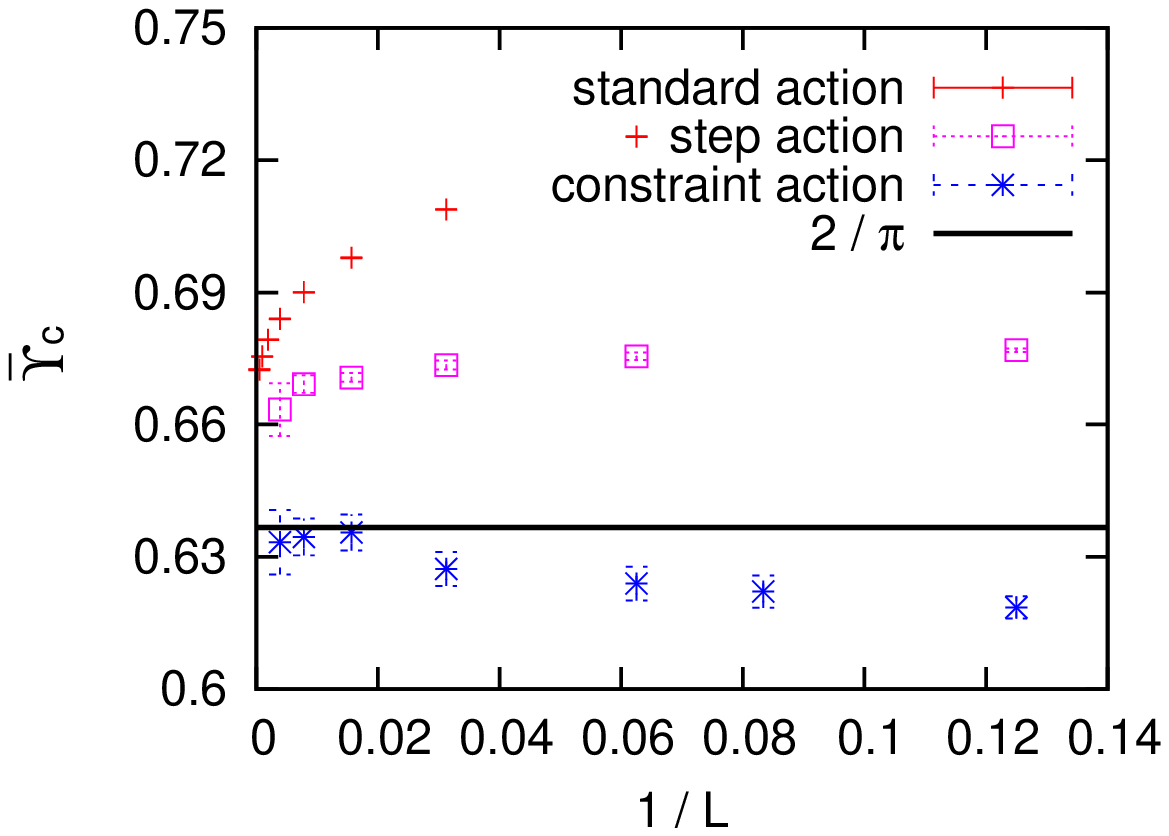}
\vspace*{-3mm}
\caption{For the constraint action,
the dimensionless helicity modulus $\bar \Upsilon$ approaches
a jump near $\delta_{\rm c}$ for increasing volume (on the left).
At $\delta_{\rm c}$ it reproduces very well the BKT value $2 / \pi$ 
(on the right).}
\vspace*{-4mm}
\label{helifig}
\end{figure}

For that case, we further measured the dimensionless helicity modulus
\be
\bar \Upsilon = - \partial^{2}_{\alpha} \ln p(\alpha )
|_{\alpha = 0} \ ,
\ee
where $\alpha$ is a twist angle in the boundary conditions.
The theoretical prediction at a BKT transition is 
$\bar \Upsilon_{\rm c} = 2 / \pi$ \cite{NelKos77}.
We simulated with {\em dynamical boundary conditions} and determined
$\bar \Upsilon$ from the curvature in the histogram for the $\alpha$ 
values \cite{XYtopactYc}. Figure \ref{helifig} on the left shows
that for increasing volume a jump down to zero is approximated at 
$\delta \gtrsim \delta_{\rm c}$, as expected. The plot on the right shows
our results at fixed $\delta_{\rm c}$, in various volumes, which we denote
as $\bar \Upsilon_{\rm c}$. Earlier studies dealt with the standard 
action \cite{MHas} and the ``step action'' \cite{OlsHol}. Those 
data are reproduced in our plot, but they confirmed the theoretical
value only with a large $L$ extrapolation. The standard action
(step action) result at $L=2048$ ($L=256$) was still 
5.6 \% (4.1 \%) off, whereas the constraint action results
for $L \geq 64$ match the prediction within the errors, and
even down to $L=8$ they deviate only by 2.8 \%.\footnote{If 
all these measurements were at a massless point, one would expect 
a universal curve $\bar \Upsilon_{\rm c}(L)$. However, since one uses 
the parameters which would be critical in infinite volume, the
correlation length is actually finite, and one obtains mixed
finite size and lattice spacing artifacts, as usual.}

\begin{figure}[h!]
\vspace*{-3mm}
\begin{center}
\includegraphics[angle=270,width=.49\linewidth]{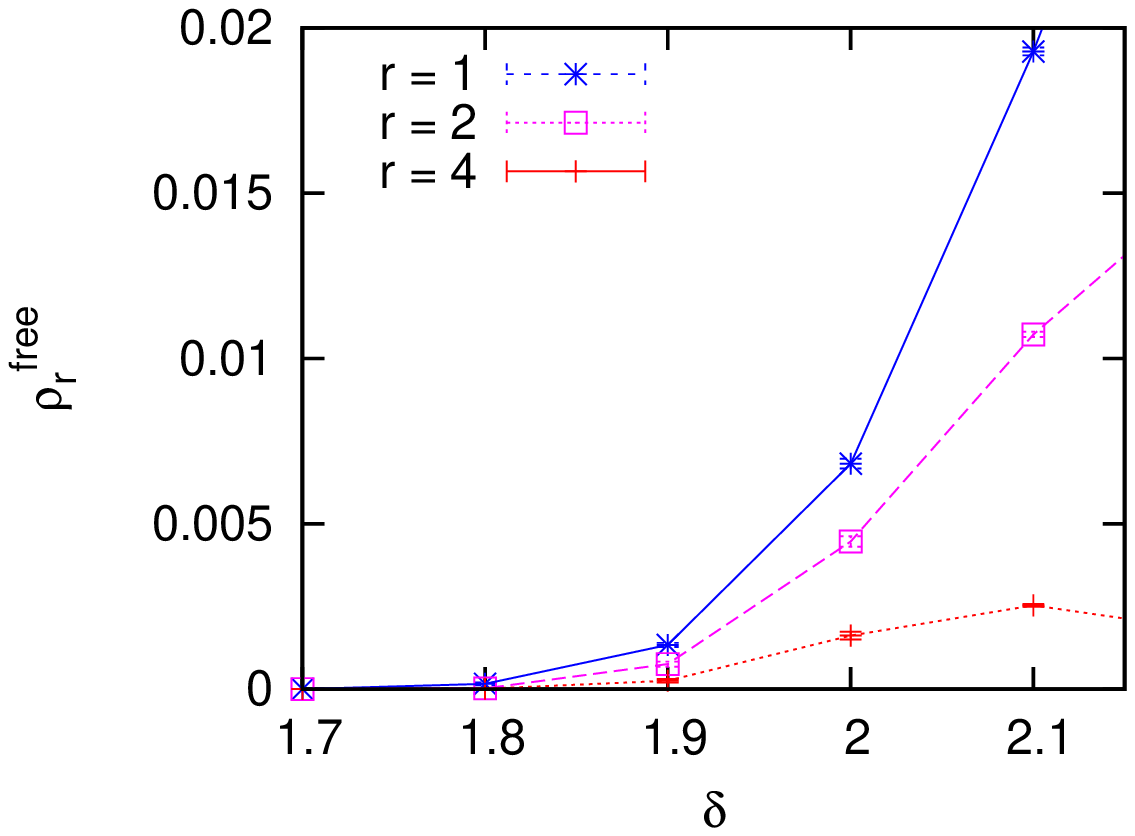}
\vspace*{-6mm}
\includegraphics[angle=270,width=.49\linewidth]{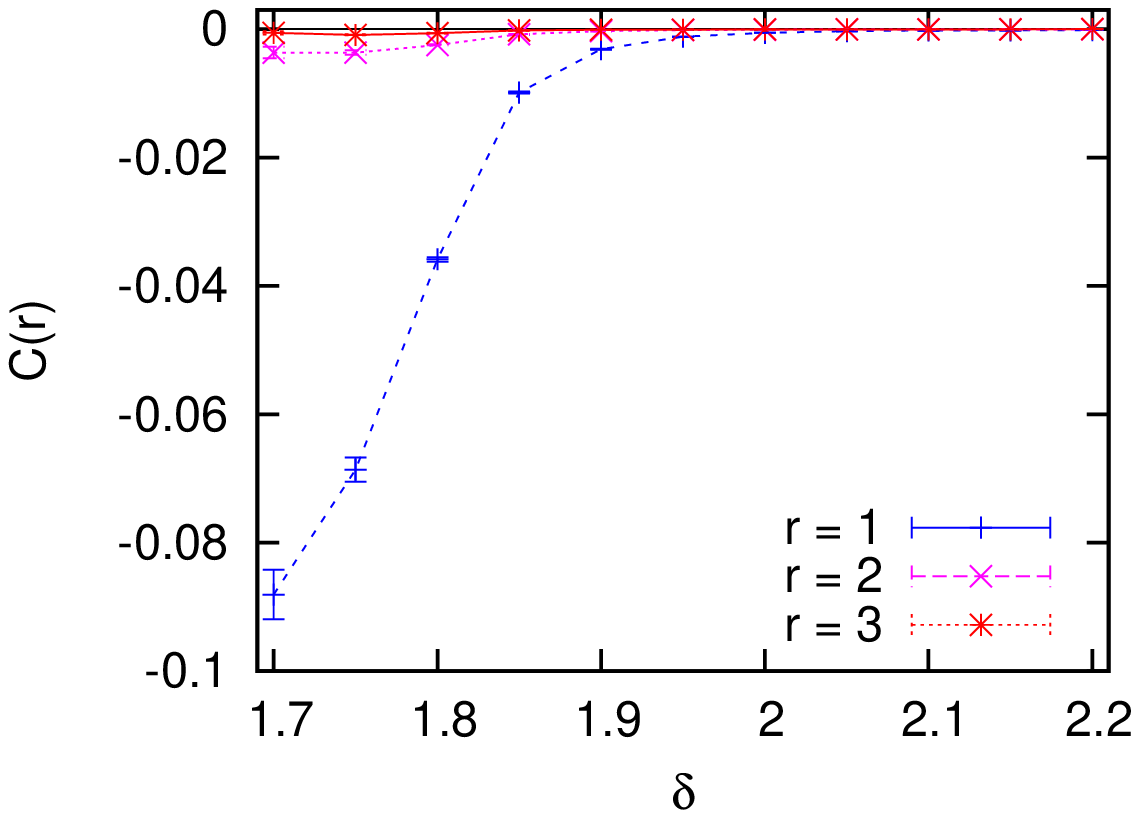}
\end{center}
\vspace*{-1.5mm}
\caption{At $\delta \gtrsim \delta_{\rm c}$  the density of 
``free vortices'' (without opposite partner within distance $r$) 
rises (left), while the vorticity anti-correlation fades
away (right). Both support the BKT (un)binding mechanism.}
\vspace*{-3mm}
\label{unbindfig}
\end{figure}

Finally we verified \cite{XYtopactYc} that the (un)binding mechanism 
of vortex--anti-vortex pairs is still valid for this transition,
even in the absence of any action cost for free vortices (and
anti-vortices). Figure~\ref{unbindfig} (left) shows the
densities of ``free vortices'', defined as vortices which do not
have an anti-vortex partner (or v.v.) within a distance of 
$r=1,\, 2$ or $4$ lattice spacings. A significant density sets in
as we increase $\delta$ somewhat above $\delta_{\rm c}$. Figure 
\ref{unbindfig} (right) shows the ``vorticity correlation function''
\be  
\vspace*{-1mm}
C(r) = \langle v_{\Box ,\, x} v_{\Box ,\, x+r} \rangle
\vert_{|v_{\Box ,\, x}|=1} \ .
\ee
In particular for $r=1$ --- nearest neighbour pairs --- 
we see a significant anti-correlation up to $\delta \approx 
\delta_{\rm c}$, which fades away as we increase $\delta$.
Along with measurements for the ``optimal pair formation'' 
(based on {\em simulated annealing}), we obtained compelling evidence 
that the (un)binding mechanism does indeed still drive the BKT 
transition, even for the constraint action \cite{XYtopactYc}.

We return to the phase diagram in Figure \ref{phasediaSSF} (on the 
left) and focus on the pure $Q$ suppressing action (which corresponds
to $\delta = \pi$). Initially we expected the phase transition
line to end somewhere on this upper axis at finite $\lambda$.
However, the simulation results match
$\xi (\lambda ) \propto \exp (0.729 \, \lambda)$, 
hence the endpoint is located at $\lambda_{\rm c} = + \infty$
(as depicted in the phase diagram) \cite{XYtopact}.
In this limit the vortices and anti-vortices are completely
eliminated, so one may question whether this can still be a BKT 
transition. Indeed, the numerical result for the SSF
\cite{XYtopact}, $\sigma (2, u=6)_{\rm num} = 9.47(1)$, differs from 
the BKT value $\sigma (2, u=6)_{\rm BKT} = 11.53$ (provided by J.\ Balog).
Therefore the transition at this endpoint is {\em not} of the BKT
type; it belongs to another class, which has apparently been
overlooked in the (tremendous) literature on this model.

\section{Conclusions}

\vspace*{-2mm}

Topological lattice actions do not have a classical limit,
nor a perturbative expansion, but in the O(N) models studied
here they do have the correct quantum continuum limit.
This underscores the enormous power of universality.
It even captures the {\em rotor} as a quantum mechanical model 
($d=1$), though with unusual linear lattice artifacts. 

In the {\em 2d O(3) model} 
the constraint action violates the Schwarz inequality,
but it has an excellent scaling behaviour, which can be 
further improved by a combination with a fine-tuned standard 
coupling constant \cite{drastic}. The term
$\chi_{t} \xi^{2}$ diverges logarithmically in the continuum
limit, but the correlator of the topological charge density,
$\langle q_{x} q_{y} \rangle \, $, remains finite for $x \neq y$.
This enabled precise studies of $\theta$ vacuum effects \cite{theta}.
%which demonstrated that the $\theta$ vacuum angle does 
% not renormalise to zero \cite{theta}.

In the {\em 2d XY model,} the $(\delta , \lambda)$ phase diagram
has a BKT transition line at finite $\lambda$ (at least up to 
$\lambda =4$). This was shown by measuring the SSF and
the critical exponent $\eta$ \cite{XYtopact}.
For the constraint action, the helicity modulus approaches
a jump near $\delta_{\rm c}$ for increasing volume. At 
$\delta_{\rm c}$ it reproduces %for the first time 
directly the BKT value $\bar \Upsilon_{\rm c} = 2/\pi$, which constitutes
one of the best numerical evidences for a BKT transition. 
It is still driven by the vortex--anti-vortex pair (un)binding 
mechanism, now as a purely combinatorial effect, since there is 
no Boltzmann factor involved. 

On the other hand, without a constraint angle, a new type of 
transition is attained at vortex suppression $\lambda \to \infty$; 
this endpoint of the transition line in the $(\delta , \lambda)$ 
phase diagram remains to be explored.

\end{document}